\documentclass[10pt]{article}
\usepackage{amssymb,amsmath,amsthm,fullpage,hyperref}

\newtheorem{theorem}{Theorem}

\theoremstyle{remark}
\newtheorem{assumption}{Assumption}

\begin{document}

\begin{figure}
\begin{flushright}
SNUST 080902
\end{flushright}
\end{figure}

\title{Tree level spontaneous R-symmetry breaking\\
in O'Raifeartaigh models}
\author{Zheng Sun\\
\normalsize\textit{Center for Theoretical Physics, Seoul National University, Seoul 151-747, Korea}\\
\normalsize\textit{E-mail:} \texttt{zsun@phya.snu.ac.kr}}
\date{}
\maketitle

\begin{abstract}
We show that in O'Raifeartaigh models of spontaneous supersymmetry breaking, R-symmetries can be broken by non-zero values of fields at tree level, rather than by vacuum expectation values of pseudomoduli at loop level. As a complement of the recent result by Shih, we show that there must be a field in the theory with R-charge different from zero and two in order for R-symmetry breaking to occur, no matter whether the breaking happens at tree or loop level. We review the example by CDFM, and construct two types of tree level R-symmetry breaking models with a wide range of parameters and free of runaway problem. And the R-symmetry is broken everywhere on the pseudomoduli space in these models. This provides a rich set of candidates for SUSY model building and phenomenology.
\end{abstract}

\section{Introduction}

O'Raifeartaigh models of spontaneous supersymmetry breaking \cite{O'Raifeartaigh:1975pr, Intriligator:2007cp} have recently received much interest in low-energy SUSY model building. In these studies, R-symmetries play important roles due to its relation to SUSY breaking \cite{Affleck:1984xz, Nelson:1993nf}, known as Nelson-Seiberg theorem. For a generic model without fine tuning, a $\operatorname{U}(1)$ R-symmetry for the superpotential is a necessary and sufficient condition for SUSY breaking. The R-symmetry needs to be broken to have non-zero Majorana gaugino masses. In many O'Raifeartaigh's models considered to date, the R-symmetry is spontaneously broken by the pseudomodulus which exists in any non-SUSY vacuum from Wess-Zumino models with minimal K\"ahler potential \cite{Ray:2006wk, Sun:2008nh}. To have this happen, the one-loop Coleman-Weinberg potential \cite{Coleman:1973jx} has to stabilize the pseudomodulus at some non-zero value, which requires that there must be a field in the theory with R-charge different from $0$ or $2$. This result of R-charge assignment is derived recently by Shih \cite{Shih:2007av} which is based on the following assumptions:

\begin{assumption} \label{as:1-10}
The R-symmetry is broken by the vacuum expectation value of the pseudomodulus. Coleman-Weinberg potential must give a negative mass to the pseudomodulus in order to stabilize it at some non-zero value.
\end{assumption}

\begin{assumption} \label{as:1-20}
Other fields all have zero vacuum expectation values.
\end{assumption}

Although most models studied up to date satisfy these assumptions, one exception has been observed in \cite{Carpenter:2008wi}. It violates both assumptions here: Some fields other than the pseudomoduli acquire non-zero vacuum expectation values at tree level and break the R-symmetry. The purpose of this paper is to investigate such models in general with tree level spontaneous R-symmetry breaking. We find these models share the same R-charge assignment property of Shih's result for models with one-loop R-symmetry breaking: there must be a field in the theory with R-charge different from $0$ or $2$. This serves as a complement of Shih's result: The same requirement for R-charge assignment has to be satisfied, no matter whether the R-symmetry is broken at tree or loop level.

The model of \cite{Carpenter:2008wi} (CDFM model) requires an extra $Z_2$ symmetry to avoid the existence of runaway which is very common in O'Raifeartaigh models \cite{Ferretti:2007ec}. The role of the extra symmetry can also be played by more complicated R-charge assignment. We propose two of such models which also have tree level R-symmetry breaking but do not need extra symmetries. The R-symmetry is broken everywhere on the pseudomoduli space, which makes a clear distinction between these models and Shih's type. All these models have wide ranges of parameters, which provide a rich set of candidates for the study of model building and phenomenology.

The outline of the paper is as follows. In section 2 we point out that it is possible to dissatisfy assumption \ref{as:1-20}: The vacuum does not need to coincide the R-invariant point. In section 3 we prove the R-charge assignment requirement for tree level R-symmetry breaking: Some field with R-charge different from $0$ or $2$ is required. In section 4 we discuss the problem of runaway and the way to avoid it. In section 5 we review the model from \cite{Carpenter:2008wi}. In section 6 we propose two types of tree level R-symmetry breaking models without the need of extra symmetries to avoid runaway.

\section{Vacuum expectation values of fields}

One fact which is often overlooked is that the field values at the vacuum are not necessarily zero. Although it is convenient to make a field redefinition so that the vacuum is set at the origin, such redefinition, which involves translations, also moves the invariant point of the R-symmetry. Then the origin after the redefinition is not necessarily R-invariant. Alternatively, one can keep the R-invariance of the origin of the field space so that the R-symmetry transformation can always be written as a simple rotation around the origin. But then vacuum expectation values of fields are not necessarily zero. So it is possible to break the R-symmetry at tree level if these fields with non-zero vacuum expectation values have non-zero R-charges.

One related result is that not every field needs an explicit mass term to be stabilized. Mass terms can be generated by the non-zero vacuum expectation values of fields. So the superpotential $W$ may have only linear and cubic terms for some fields. Realizing this gives more freedom on model building: Not every field with R-charge $q$ needs a partner with R-charge $2-q$ to form an explicit mass term.

\section{R-charge assignment condition}

We are to prove the requirement for R-charge assignment for tree level R-symmetry breaking. If there are only R-charge $2$ fields $X_i$, $i=1, \ldots, d_X$ and R-charge $0$ fields $Y_J$, $J=1, \ldots, d_Y$, the superpotential can be written as
\begin{equation} \label{eq:3-10}
W = \sum_i X_i f_i(Y_J)
\end{equation}
where $f_i$ are polynomial functions of R-neutral fields $Y_J$. If we constrain our consideration to renormalizable theories, $f_i$ can only have up to quadratic terms. But our following proof also applies to non-renormalizable superpotentials. The field strength is
\begin{equation}
\partial_i W = f_i(Y_J), \quad \partial_I W = \sum_i X_i \partial_I f_i(Y_J) \ .
\end{equation}
We need to minimize the scalar potential
\begin{equation}
V = \sum_i \lvert f_i(Y_J) \rvert^2 + \sum_I \lvert \sum_i X_i \partial_I f_i(Y_J) \rvert^2 \ .
\end{equation}

The second term in the expression of $V$ must vanish at the vacuum. If it does not vanish, i.e.\ there is a local minimum at $(X_i^{(0)}, Y_J^{(0)})$ so that
\begin{equation}
V^{(0)} = \sum_i \lvert f_i(Y_J^{(0)}) \rvert^2 + \sum_I \lvert \sum_i X_i^{(0)} \partial_I f_i(Y_J^{(0)}) \rvert^2 = V_0 + V_1, \quad V_1 > 0 \ ,
\end{equation}
$X_i^{(0)}$ must not be all zero, otherwise $V_1$ will vanish. Consider the field subspace
\begin{equation} \label{eq:3-20}
(X_i, Y_J)(c) = (c X_i^{(0)}, Y_J^{(0)}), \quad c \in \mathbb{C} \ .
\end{equation}
On this subspace we have
\begin{equation}
V(c) = V(c X_i^{(0)}, Y_J^{(0)}) = V_0 + \lvert c \rvert^2 V_1
\end{equation}
and $V(1) = V^{(0)}$. The derivative $\partial_c V = c^* V_1$ is non-zero at $c=1$, which contradict the assumption that $(X_i^{(0)}, Y_J^{(0)})$ is a metastable vacuum. So we see $V_1$ must vanish at the vacuum.

There are two ways to make $V_1 = 0$: If $X_i = 0$, they do not break the R-symmetry. R-neutral fields $Y_J$, even if they have non-zero vacuum expactation values, also do not break the R-symmetry. So the R-symmetry is preserved. If $X_i \ne 0$, $\partial_I f_i(Y_J)$ take some special values so that the combination to $V_1$ vanishes. Then some combination of $X_i$ is a pseudomodulus. One way to see this is to define a subspace like \eqref{eq:3-20}, then $c$ labels the flat direction. The R-symmetry is broken on the pseudomoduli space except at the origin. One needs to do one-loop computation to determine where the pseudomodulous is stabilized. So it falls into the scope of Shih's result \cite{Shih:2007av}. In this case whether the R-symmetry is broken or not is undetermined at tree level.

When there are fields with R-charges other than $2$ and $0$, one may redefine fields following the step of the proof of Nelson-Seiberg theorem \cite{Nelson:1993nf} and make the superpotential has the form of \eqref{eq:3-10}. But the redefinition is singular at the origin, and also makes the K\"ahler potential non-minimal. So the scalar potential is more complicated and we can't make the above argument.

Another way to see R-charge $2$ fields can not break the R-symmetry at tree level is to consider the complexification of the symmetry group as done in \cite{Sun:2008nh}. The R-symmetry rotates fields as well as the superpotential, and also the SUSY breaking field strength:
\begin{equation} \label{eq:3-30}
\partial_i W \rightarrow e^{i (2 - q_i) \alpha} \partial_i W, \quad \alpha \in \mathbb{R} \ .
\end{equation}
When the R-charge $q_i = 2$, $\partial_i W$ is invariant under the R-symmetry even if $\alpha$ is taken to be complex. For R-charge $q_i = 0$ fields, the corresponding field strength components vanish according to the first step of the previous proof. So the scalar potential $V$ is also invariant under the complexified R-symmetry. If some R-charge $2$ fields have non-zero values at the vacuum, the complexified R-symmetry makes a whole complex plane as a pseudomoduli space. So one needs to consider one-loop correction to stabilize the pseudomodulus. At tree level one can not determine whether R-symmetry is broken or not.

We have proved that the R-symmetry can not be broken (or it is undetermined) at tree level if there are only R-charge $0$ and $2$ fields. Combining Shih's result \cite{Shih:2007av} with our proof, we have the conclusion:
\begin{theorem}
There must be a field in the theory with R-charge different from $0$ and $2$ in order for R-symmetry breaking to occur, no matter whether the breaking happens at tree or loop level.
\end{theorem}
On the other hand, one can build tree level spontaneous R-symmetry breaking models which contain fields with R-charges other than $0$ and $2$. One apparent distinction between these models and Shih's type is that whether there is any R-symmetry preserving point on the pseudomoduli space. In the tree level breaking case, the breaking happens everywhere on the pseudomoduli space. So whatever one-loop computation is, the R-symmetry is always broken. We are to show such examples in later sections.

\section{The problem of runaway}

Before building any tree level R-symmetry breaking model, we would like to look into a common feature in O'Raifeartaigh models: runaway directions, where SUSY is asymptotically restored as some fields approaching infinity. Having runaway directions may not always be a problem, since the matastable vacuum may still have a long lifetime against quantum tunneling to runaway, or quantum corrections may stabilize the fields at finite values. Even though, it is still worthwhile to know the condition of avoiding runaway which gives an alternative way of model building. Runaway directions are usually related to R-symmetries. Here we are to summarize the result from \cite{Carpenter:2008wi, Ferretti:2007ec}. The field strength transforms under the R-symmetry as \eqref{eq:3-30}. We categorize the SUSY equations according to their R-charges:
\begin{equation}
\partial_i W = 0, \quad
\begin{cases}
q_i > 2, \quad \operatorname{R}(\partial_i W)< 0 \\
q_i = 2, \quad \operatorname{R}(\partial_i W)= 0 \\
q_i < 2, \quad \operatorname{R}(\partial_i W)> 0
\end{cases} \ .
\end{equation}
If all equations can be satisfied simultaneously, one gets a SUSY vacuum, otherwise the model has only non-SUSY vacua. If one can just satisfy the $q_i \ge 2$ equations, the complexified R-transformation $z_i \rightarrow z_i e^{q_i \alpha}$ does not affect these equations. Taking $\alpha \rightarrow -\infty$, the $q_i < 2$ equations are also satisfied at the limit, so it is a SUSY runaway direction. Similarly, if one can satisfy the $q_i \le 2$ equations, then $\alpha \rightarrow \infty$ will be a SUSY runaway direction. In both cases, generically the number of equations which need to be satisfied is less than the number of variables, and the solution exists. To avoid runaway, one needs to consider some non-generic model so that these equations can not be satisfied, i.e.:

\begin{theorem}
A necessary condition to avoid runaway is that the subset of equations $\partial_i W = 0$ with $q_i \ge 2$ can not be satisfied, and the subset with $q_i \le 2$ also can not be satisfied.
\end{theorem}

We will see examples with or without runaway in the following sections. One should notice that here we only give a necessary condition. It is a sufficient condition to avoid only the specific type of runaway which is related to the R-symmetry. Even if they are satisfied, one still needs to be careful about the existence of other types of runaway.

\section{Review of CDFM model}

An example of tree level R-symmetry breaking, CDFM model, has been observed in \cite{Carpenter:2008wi}. Here we are to review its vacuum structure which is very similar to our models in the next section. The original model has five chiral fields and the superpotential
\begin{equation}
W = \lambda z_1 (z_4 z_5 - m^2) + \mu z_2 z_4 + \nu z_3 z_5 + \sigma z_5^3 \ .
\end{equation}
The R-charge assignment for $z_1, \ldots, z_5$ is:
\begin{equation}
q_1 = 2, \quad q_2 = 8/3, \quad q_3 = 4/3, \quad q_4 = -2/3, \quad q_5 = 2/3 \ .
\end{equation}
The components of the SUSY breaking field strength
\begin{equation} \label{eq:5-10}
\begin{gathered}
\partial_1 W = \lambda (z_4 z_5 - m^2), \quad \partial_2 W = \mu z_4, \quad \partial_3 W = \nu z_5, \\
\partial_4 W = \lambda z_1 z_5 + \mu z_2, \quad \partial_5 W = \lambda z_1 z_4 + \nu z_3 + 3 \sigma z_5^2
\end{gathered}
\end{equation}
can not be set to zero simultaneously, so there is no SUSY vacuum for this model. We need to minimize the potential
\begin{equation}
V = \lvert \lambda \rvert^2 \lvert z_4 z_5 - m^2 \rvert^2 + \lvert \mu \rvert^2 \lvert z_4 \rvert^2 + \lvert \nu \rvert^2 \lvert z_5 \rvert^2 + \lvert \lambda z_1 z_5 + \mu z_2 \rvert^2 + \lvert \lambda z_1 z_4 + \nu z_3 + 3 \sigma z_5^2 \rvert^2 \ .
\end{equation}
By field redefinition by phases, all coefficients can be made real and non-negative. Assuming they are positive and satisfy
\begin{equation}
\mu \nu < \lambda^2 m^2 \ ,
\end{equation}
the non-SUSY vacuum satisfies
\begin{equation}
\lvert \mu z_4 \rvert = \lvert \nu z_5 \rvert, \quad z_4 z_5 = m^2 - \frac{\mu \nu}{\lambda^2}, \quad \lambda z_1 z_5 + \mu z_2 = 0, \quad \lambda z_1 z_4 + \nu z_3 + 3 \sigma z_5^2 = 0 \ .
\end{equation}
There is also another extremum of $V$ where $z_1$ is the pseudomodulus and $z_2, \ldots, z_5$ are set to zero, but it has higher $V$. The solution we provide above is actually the global minimum of the potential. $z_4, z_5$ have non-zero vacuum expectation values
\begin{equation}
z_4 = \nu r e^{i \theta}, \quad z_5 = \mu r e^{-i \theta}, \quad \theta \in \mathbb{R}, \quad r = \sqrt{\frac{m^2}{\mu \nu} - \frac{1}{\lambda^2}} \ .
\end{equation}
The R-symmetry is spontaneously broken by the non-zero $r$ and the R-axion is labeled by $\theta$. The pseudomodulus from the theorem of \cite{Ray:2006wk, Sun:2008nh} is
\begin{equation}
z'_1 = A^{-1} (z_1 - \lambda r z_2 e^{-i \theta} - \lambda r z_3 e^{i \theta}), \quad A = \sqrt{\frac{2 \lambda^2 m^2}{\mu \nu}  - 1}
\end{equation}
which should be viewed as a linear redefinition of $z_1, z_2, z_3$ with $\theta$ fixed. So the total pseudomoduli space is of real dimension $3$: $2$ from  the theorem of \cite{Ray:2006wk, Sun:2008nh} and $1$ from the R-axion. The R-symmetry is spontaneously broken everywhere on the 3-dimensional pseudomoduli space. Loop corrections will further stabilize the value of $z'_1$.

The first three equations of \eqref{eq:5-10} meet the necessary condition of avoiding runaway which we discussed in the previous section. Also there is no other type of runaway direction in this model. However, as already pointed out in \cite{Carpenter:2008wi}. The R-symmetry allows another term for $W$:
\begin{equation}
\delta W = \epsilon z_3^2 z_4
\end{equation}
which introduces the problem of runaway:
\begin{equation}
\begin{gathered}
z_1 = (\frac{\nu^2 m^2}{2 \lambda \epsilon}-\frac{3 \sigma m^4}{\lambda}) e^{3 \alpha}, \quad z_2 = (\frac{3 \sigma m^6}{\mu}-\frac{3 \nu^2 m^4}{4 \mu \epsilon}) e^{4 \alpha}, \\
z_3 = -\frac{\nu m^2}{2 \epsilon}e^{2\alpha}, \quad z_4 = e^{-\alpha}, \quad z_5 = m^2 e^\alpha, \quad \alpha \rightarrow \infty \ .
\end{gathered}
\end{equation}
This flaw can be fixed by adding a field $z_6$ with R-charge $q_6 = 2/3$, and introducing an extra $\operatorname{Z}_2$ symmetry under which $z_1, z_6$ are even and other fields are odd. The superpotential is taken to be
\begin{equation}
W = \lambda z_1 (z_4 z_5 - m^2) + \mu z_2 z_4 + \nu z_3 z_5 + \tau z_5^2 z_6 + \sigma z_6^3 \ .
\end{equation}
No other renormalizable term is allowed by the symmetries. So the problem of runaway is avoided. This model has similar vacuum spectrum as the previous one: SUSY and the R-symmetry are spontaneously broken everywhere on the 3-dimensional pseudomoduli space.

\section{Models without extra symmetries}

Although extra symmetries in the hidden sector like the $\operatorname{Z}_2$ in CDFM model may not be a problem for realistic model building, we would like to seek models with similar vacuum spectrum which do not depend on the extra symmetry. This may give more freedom for model building. The purpose of the extra symmetry is to prevent extra terms which may introduce the problem of runaway. One may explore more complicated R-charge assignment to serve the same purpose, as we are to do in this section.

We propose two types of models which have $7$ chiral fields and the superpotential
\begin{equation}
W = W_0 + W_1 = \lambda z_1 (z_4 z_5 - m^2) + \mu z_2 z_4 + \nu z_3 z_5 + W_1(z_4, \ldots, z_7) \ .
\end{equation}
The R-charge assignment is
\begin{equation} \label{eq:6-10}
q_1 = 2, \quad q_2 = 2+q, \quad q_3 = 2-q, \quad q_4 = -q, \quad q_5 = q, \quad q_6 = 2-2q, \quad q_7 = 3q \ .
\end{equation}
The form of $W_0$ ensures SUSY breaking and R-symmetry breaking by granting vacuum expectation values to $z_4$ and $z_5$ in a similar way as CDFM model does. And $W_1$ contains all other renormalizable terms which are allowed by the R-symmetry. The two types of models, which have different values of $q$, are named after the form of the last term of $W_1$ which uniquely determines the R-charges of all fields.

\begin{enumerate}
\item The $M z^2$ model: $q = 1/3$. The extra terms in $W_1$ are
\begin{equation}
W_1 = a z_5^2 z_6 + b z_4 z_6 z_7 + M z_7^2 \ .
\end{equation}
And the R-charge assignment \eqref{eq:6-10} is
\begin{equation}
q_1 = 2, \quad q_2 = 7/3, \quad q_3 = 5/3, \quad q_4 = -1/3, \quad q_5 = 1/3, \quad q_6 = 4/3, \quad q_7 = 1 \ .
\end{equation}
\item The $\sigma z^3$ model: $q = 2/9$. The extra terms in $W_1$ are
\begin{equation}
W_1 = a z_5^2 z_6 + b z_4 z_6 z_7 + \sigma z_7^3 \ .
\end{equation}
And the R-charge assignment \eqref{eq:6-10} is
\begin{equation}
q_1 = 2, \quad q_2 = 20/9, \quad q_3 = 16/9, \quad q_4 = -2/9, \quad q_5 = 2/9, \quad q_6 = 14/9, \quad q_7 = 2/3 \ .
\end{equation}\end{enumerate}

In both models, $W_1$ does not depend on $z_1, \ldots, z_3$. And $W_0$ have the same form as part of the superpotential in CDFM model. So the following solution also has many similarities as the solution of CDFM model. The SUSY breaking field strength is
\begin{equation}
\begin{gathered}
\partial_1 W = \lambda (z_4 z_5 - m^2), \quad \partial_2 W = \mu z_4, \quad \partial_3 W = \nu z_5, \\
\partial_4 W = \lambda z_1 z_5 + \mu z_2 + b z_6 z_7, \quad \partial_5 W = \lambda z_1 z_4 + \nu z_3 + 2 a z_5 z_6, \\
\partial_6 W = a z_5^2 + b z_4 z_7, \quad \partial_7 W =
\begin{cases}
b z_4 z_6 + 2 M z_7, \quad \text{for the } M z^2 \text{ Model} \\
b z_4 z_6 + 3 \sigma z_7^2, \quad \text{for the } \sigma z^3 \text{ Model}
\end{cases} \ .
\end{gathered}
\end{equation}
The first three components can not be set to zero simultaneously, so there is no SUSY vacuum for this model. We need to minimize the potential
\begin{equation}
V = \lvert \lambda \rvert^2 \lvert z_4 z_5 - m^2 \rvert^2 + \lvert \mu \rvert^2 \lvert z_4 \rvert^2 + \lvert \nu \rvert^2 \lvert z_5 \rvert^2 + \ldots \ .
\end{equation}
By field redefinition by phases, all coefficients can be made real and non-negative. Assuming they are positive and satisfy
\begin{equation} \label{eq:6-20}
\mu \nu < \lambda^2 m^2 \ ,
\end{equation}
the non-SUSY vacuum satisfies
\begin{equation} \label{eq:6-30}
\lvert \mu z_4 \rvert = \lvert \nu z_5 \rvert, \quad z_4 z_5 = m^2 - \frac{\mu \nu}{\lambda^2}, \quad \partial_i W = 0, \quad i = 4, \ldots, 7 \ .
\end{equation}
There is also another extremum of $V$ where $z_1$ is the pseudomodulus and all other fields are set to zero. But it has higher $V$. The solution we provide above is actually the global minimum of the potential. Our class of models also have no runaway direction. $z_4, z_5$ have non-zero vacuum expectation values
\begin{equation} \label{eq:6-40}
z_4 = \nu r e^{i \theta}, \quad z_5 = \mu r e^{-i \theta}, \quad \theta \in \mathbb{R}, \quad r = \sqrt{\frac{m^2}{\mu \nu} - \frac{1}{\lambda^2}} \ .
\end{equation}
The R-symmetry is spontaneously broken by the non-zero $r$ and the R-axion is labeled by $\theta$. The pseudomodulus from the theorem of \cite{Ray:2006wk, Sun:2008nh} is
\begin{equation} \label{eq:6-50}
z'_1 = A^{-1} (z_1 - \lambda r z_2 e^{-i \theta} - \lambda r z_3 e^{i \theta}), \quad A = \sqrt{\frac{2 \lambda^2 m^2}{\mu \nu}  - 1}
\end{equation}
which should be viewed as a linear redefinition of $z_1, z_2, z_3$ with $\theta$ fixed. So the total degeneracy space is of real dimension $3$: $2$ from the theorem of \cite{Ray:2006wk, Sun:2008nh} and $1$ from the R-axion. The R-symmetry is spontaneously broken everywhere on the 3-dimensional pseudomoduli space. Loop corrections will further stabilize the value of $z'_1$. All other field values can be expressed in terms of the R-axion $\theta$ by solving \eqref{eq:6-30}. For the $M z^2$ model, the vacuum can be described by \eqref{eq:6-40}, \eqref{eq:6-50} and
\begin{equation}
\begin{gathered}
\left.
\begin{gathered}
z'_2 = B^{-1} (\lambda r z_1 e^{-i \theta} + z_2 - \frac{2 M a^2 \mu^3}{b^2 \nu^3} r e^{-7 i \theta}) = 0, \\
z'_3 = B^{-1} (\lambda r z_1 e^{i \theta} + z_3 + \frac{4 M a^2 \mu^3}{b^2 \nu^3} r e^{-5 i \theta}) = 0,
\end{gathered}
\right\} \quad B = \frac{\lambda m}{\sqrt{\mu \nu}}, \\
z_6 = \frac{2 M a \mu^2}{b^2 \nu^2} e^{-4 i \theta}, \quad z_7 = - \frac{a \mu^2}{b \nu} r e^{-3 i \theta} \ .
\end{gathered}
\end{equation}
For the $\sigma z^3$ model, the vacuum can be described by \eqref{eq:6-40}, \eqref{eq:6-50} and
\begin{equation}
\begin{gathered}
\left.
\begin{gathered}
z'_2 = B^{-1} (\lambda r z_1 e^{-i \theta} + z_2 + \frac{3 \sigma a^3 \mu^5}{b^3 \nu^4} r^2 e^{-10 i \theta}) = 0, \\
z'_3 = B^{-1} (\lambda r z_1 e^{i \theta} + z_3 - \frac{6 \sigma a^3 \mu^5}{b^3 \nu^4} r^2 e^{-8 i \theta}) = 0,
\end{gathered}
\right\} \quad B = \frac{\lambda m}{\sqrt{\mu \nu}}, \\
z_6 = - \frac{3 \sigma a^2 \mu^4}{b^3 \nu^3} r e^{-7 i \theta}, \quad z_7 = - \frac{a \mu^2}{b \nu} r e^{-3 i \theta} \ .
\end{gathered}
\end{equation}
In both models, $z'_2, z'_3$ should be viewed as linear redefinitions of $z_1, z_2, z_3$ with $\theta$ fixed. The normalization factors $A$ and $B$ is used to make the field redefinition $(z_1, z_2, z_3) \rightarrow (z'_1, z'_2, z'_3)$ a unitary transformation so that the K\"ahler potential remains a minimal form.

These models, although have non-generic R-charge assignment, do have a wide range of parameters. All needs to be satisfied is just the condition \eqref{eq:6-20}. Coupling to one of many candidate SUSY mediation and SSM models, the possibility of tree level R-symmetry breaking opens up many interesting directions for model building and phenomenology.

\section*{Acknowledgement}
The author would like to thank Michael Dine, Soo-Jong Rey, Satoshi Yamaguchi and Sangheon Yun for helpful discussions. This work is supported by BK-21 Program, KRF-2005-084-C00003, EU FP6 Marie Curie Research \& Training Networks HPRN-CT-2006-035863.

\end{document}